\documentclass[journal]{IEEEtran}
   \usepackage[dvips]{graphicx}
 \usepackage{color}

\usepackage[cmex10]{amsmath}

\begin{document}

\title{Dispersive wave emission in dual concentric core fiber: the role of soliton-soliton collisions}

\author{~Alessandro~Tonello, Daniele~Modotto,~\IEEEmembership{Member,~IEEE,} Katarzyna~Krupa,
        Alexis~Labruy\`ere, Badr~Shalaby, Vincent~Couderc, Alain~Barth\'el\'emy, 
        Umberto~Minoni, Stefan~Wabnitz,~\IEEEmembership{Member,~IEEE,}
        and Alejandro~B.~Aceves
\thanks{A. Tonello, K. Krupa, A. Labruy\`ere, B. Shalaby, V. Couderc and
A. Barth\'el\'emy are with the Universit\'e de Limoges, XLIM, UMR CNRS 7252, 123 Av. A. Thomas, 87060 Limoges, France. B. Shalaby is also with the Physics Department, Tanta University, Tanta, Egypt. D. Modotto, U. Minoni and S. Wabnitz are with the Dipartimento di Ingegneria dell'Informazione, Universit\`a di Brescia, via Branze 38, 25123 Brescia, Italy. A. B. Aceves is with the
Department of Mathematics, Southern Methodist University, Dallas, USA.}
}

\maketitle

\begin{abstract}
Soliton-soliton collisions have a crucial role in enhancing the spectrum of dispersive waves in optical fibers and collisions among in-phase solitons lead to a dramatic enhancement of the dispersive wave power, as well as to its significant spectral reshaping. 
We obtained a simple analytical model to estimate  the spectral position, width and amplitude of the dispersive waves 
induced by a collision of two in-phase solitons. We tested our theory in the case of a dual concentric core microstructured fiber.
\end{abstract}


\IEEEpeerreviewmaketitle

\section{Introduction}
\IEEEPARstart{D}{ispersive} wave (DW) emission by intense optical  {\color{black} pulses propagating in fibers has been widely studied in the past three decades  \cite{wai86}-\cite{skr}. In recent years, dispersion profile engineering in special optical fibers enabled a dramatic increase in the efficiency of DW generation,  as well as the exploitation of new mechanisms for supercontinuum 
generation  \cite{faccio,skryabin}. DWs may be used to generate new spectral components that are useful for nonlinear spectroscopy applications: therefore it is of paramount importance to be able to predict their spectral shape, as well as their efficiency and tunability}. Optical solitons in fibers may generate DWs in the presence 
of a frequency-dependent group velocity dispersion (GVD): therefore DW generation requires the presence of higher-order dispersion (HOD) terms. 
Indeed, for a soliton pulse to leak some of its energy into a DW, it is necessary that these two objects satisfy a condition of phase matching. 
The guided mode linear propagation constant $\beta(\omega)$ is typically represented in terms of a polynomial expansion in $\omega$, which is the angular frequency shift 
from a carrier frequency $\omega_0$.  
Proper dispersion engineering of specialty fibers is crucial for tailoring the HOD terms, so that one may control both spectral position and energy of the emitted DWs for any given pumping configurations \cite{agrawal}. In particular, linear-mode coupling in a dual-core  microstructured optical fiber (MOF) is responsible for a huge frequency dependence of the supermode GVD whenever the phase velocities of the two guided modes cross at a certain wavelength \cite{yariv}. This property has recently led to the demonstration of intermodal frequency conversion using femtosecond soliton pumping \cite{cheng}, as well as to the generation of gigantic dispersive waves with
sub-nanosecond pumping conditions \cite{us}.
 
 It is known that the break-up of a sub-nanosecond pump pulse can lead to the generation of a large number 
of femtosecond solitons. In this work we take into consideration the simplest case of a collision 
of two of these solitons and we study how  soliton collisions can shape and enhance the DW emission
with a simplified analytical approach.

We shall consider the first mode of a MOF, whose strongly frequency-dependent $\beta(\omega)$ 
strongly contributes to the large resonant enhancement of DW emission during soliton collisions.  
First we shall identify the source of resonant DW emission that is activated during the finite lifetime of the inelastic two-soliton collisions \cite{Erkintalo2010,erkintalocoll}. 
{\color{black} Next we will elucidate, both numerically and analytically, the specific features of DWs generated by colliding solitons in dual-core fibers. As we shall see, the DW energy keeps a memory of the collision, by exhibiting a sharp transient in a specific spectral region, similarly to a flip-flop circuit that changes its state (here the spectral intensity) in the presence of a trigger signal (here the ultrafast pulse collision). 
Thus pulse collisions act as an ultrafast  gate for emitting dispersive pulses whose spectral shape, group speed and central wavelength 
are controlled by the fiber dispersion profile. Potential applications can be envisaged in time-resolved nonlinear spectroscopy, or as a building block of complex temporal dynamics when many of these collisions are present.
}

\section{Analytical approach}
Pulse propagation in microstructured fibers can be described in terms of the perturbed scalar nonlinear Schr\"odinger equation (NLSE)
{\color{black}
\begin{equation}
i\frac{\partial U }{\partial z}-\frac{1}{2}\kappa_{0}''\frac{\partial^2 U }{\partial t^2}+\gamma|U|^2U=-\epsilon H(U)
\label{eqq:nlse}
\end{equation}
where $U(z,t)$ is the pulse envelope, $t$ is a retarded time, $\kappa_{0}''$ is the GVD at $\omega_0$,
$\gamma$ is the nonlinear coefficient and $\epsilon$ is a small parameter, to emphasize the perturbative role of the HOD operator H}.
The dispersion relation for linear waves is $\kappa(\omega)=\beta(\omega)-\beta_0-\beta_1\omega$, where $\beta_0$ and $\beta_1$
are the linear wavenumber of the mode and the inverse of the group velocity {\color{black} at $\omega_0$ and 
$\beta(\omega)=(\omega+\omega_0)n_{\text{eff}}(\omega+\omega_0)/c$, being $n_{eff}$  the frequency dependent effective index. 
We split $\kappa(\omega)$ into a constant GVD term and an operator $\hat{H}$ for the HOD: 
$\kappa(\omega)=\omega^2\kappa_{0}''/2+\epsilon \hat{H}(\omega)$, 
where  $\hat{H}(\omega)={\cal F}[H]$ and  ${\cal F}$ denotes the Fourier transform defined as
$A(\omega)=\int_ {-\infty}^{+\infty}A(t)\exp(i\omega t)dt$ (see also Ref. \cite{akhmediev}).}   
In Eq. (\ref{eqq:nlse}) we neglect the Raman response of the fiber.
As input condition to Eq. (\ref{eqq:nlse}), let us consider a pair of solitons, say, $U_0$ and $U_1$ (for $\epsilon=0$) with different frequencies. 
As a first approximation in the calculation of DW emission, we may neglect the change of soliton parameters under the {\color{black}action of $H(U)$ \cite{koda}.  
Equation (\ref{eqq:nlse}) is written in a reference frame moving at the group velocity of the first soliton}
$U_0(t,z)=A_0(t)exp(i\kappa_{S0}z)$, where $A_0(t)=\sqrt{P_0} sech(t/T_0)$, and $\kappa_{S0}=\gamma P_0/2$ is the soliton wavenumber. 
Similarly, for a second soliton $U_1$ which is initially frequency shifted by $\Omega$ with respect to $U_0$, we may write  $U_1(t,z)=A_1(t-\kappa_{\Omega}'z)exp[i(\kappa_{S1}+\kappa(\Omega))z-i\Omega t]$, where $A_1(t)= \sqrt{P_1}  sech(t/T_1)$ and $\kappa_{S1}=\gamma P_1/2$. 
{\color{black}
In the absence of HOD ( so for $\epsilon=0$), 
$\kappa(\Omega)=\kappa_{0}''\Omega^2/2$, and the group velocity mismatch  is $\kappa_{\Omega}'=\kappa_{0}''\Omega$.
The spectrum of a relatively weak DW, say, $F_k(\omega,z)={\cal F}[f_k(t,z)]$ in the presence of a single intense soliton 
pulse $U_k(t,z)$ (with $k=0,1$) can be obtained by linearizing Eq. (\ref{eqq:nlse}) with $U(t,z)=U_k(t,z)+\epsilon f_k(t,z)$. By collecting
all terms proportional to $\epsilon$},
\begin{equation}
i\frac{\partial F_{k} }{\partial z}+\kappa(\omega)F_{k} =S_k(\omega,z)
\label{eqq:diflin}
\end{equation}
The forcing terms corresponding to the two individual solitons are $S_{0,1}=-\hat{H}(\omega){\cal F}[U_{0,1}(t,z)](\omega)$.
The solution for the function $F_1$ with zero input condition reads as
\begin{equation}
F_1=\frac{\hat{H}(\omega)\hat{A}_1(\omega-\Omega)exp[i\kappa(\omega)z] }{\Delta\kappa_1(\omega)}(exp[i\Delta\kappa_1(\omega) z]-1)
\label{eqq:s1}
\end{equation}
where $\Delta\kappa_1(\omega)=\kappa_{S1}+\kappa(\Omega)+(\omega-\Omega)\kappa_{\Omega}'-\kappa(\omega)$.
{\color{black} The solution for $F_0$ can be obtained by replacing $\hat{A}_1(\omega-\Omega)$
with $\hat{A}_0(\omega)$ and $\Delta\kappa_1$ with  $\Delta\kappa_0(\omega)=\kappa_{S0}-\kappa(\omega)$. 
Resonances may appear in Eq. (\ref{eqq:s1}) at frequencies $\omega$ satisfying $\Delta\kappa_1(\omega)=0$. 
 More details of this procedure for the case of a single soliton and in presence of third and fourth order dispersion may be found in \cite{akhmediev}}.
While the sources $S_0$ and $S_1$ remain active 
for the entire length of the fiber or as long as the corresponding solitons exist, 
here we show that under 
the right conditions, collisions give the dominant contribution of DW.
Indeed,  whenever the two solitons collide in-phase they can temporarily merge into 
a single and much brighter pulse: such merging occurs for a short distance, say, $L_W\propto\Omega^{-1}$. For two solitons of similar amplitude (and duration), the ``fusion pulse'', or flash, may have up to four times higher peak power and, more importantly, halved temporal duration with respect to each of the colliding solitons \cite{Erkintalo2010,gordon2}. 
To estimate the associated DW spectrum, we may simply model the flash, which is localized both in time and space, as a hyperbolic secant in the time-domain multiplied by a Gaussian profile
in the spatial domain. We set $U_F(t,z)=A_F(t-\kappa_{\Omega_F}'z)exp[i(\kappa_{F}+\kappa(\Omega_F))z-i\Omega_F t]exp(-(z-z_0)^2/L_W^2)$, where $\kappa_{F}$ is a power-dependent wavenumber and $\Omega_F$ is the frequency shift of the flash. 
Here the Gaussian function is used to localize the flash around the
collision center $z_0$ \cite{kim}. 
During its ephemeral fate limited by $L_W$, the flash can be approximated by a soliton and therefore
can act as a source of DWs.    
The corresponding perturbation $F_F(\omega)$ also obeys
Eq. (\ref{eqq:diflin}), where the forcing term $S_k$ is replaced by $S_F(\omega,z)=-\hat{H}(\omega){\cal F}[U_F(t,z)](\omega)$, that is only activated over a short distance $L_W$ centered at $z_0$.
Since the soliton is $U_0(t)\propto \eta sech(\eta t) $, the corresponding source term is $S_0\propto  sech(\pi \omega/2\eta)$. Therefore if the temporal duration of the flash $U_F(t)$ is reduced by a factor $a$ with respect to that of the soliton, then the spectrum is
$S_F(\omega,z)\propto sech(\pi \omega/(2a\eta))$. Resonant enhancement of the DWs occurs at the spectral tails of pumping pulses and one obtains, as we shall see, that a twofold time narrowing of the pumping pulse may lead to up to a several orders of magnitude enhancement of the DW amplitude. 
Therefore, in spite of the fact that the forcing induced by $S_F(\omega,z)$ only appears for a very limited length around the soliton-soliton collision event, the total amount of DW which is generated
by a flash source may still be substantially larger than the DW which is continuously emitted by each individual soliton all along the fiber.  
In general, Eq. (\ref{eqq:diflin}) may be numerically solved for $F_F(\omega,z)$ by considering the exact shape of the fusion pulse $U_F(t,z)$. With our simplified Gaussian {\it ansatz} for the spatial dependence of the flash and for in-phase solitons of nearly equal amplitude, the analytically estimated DW spectrum reads as 
{\color{black}
\begin{equation}
F_F(\omega,z)=\frac{\sqrt{\pi}}{2}\hat{H}(\omega)\hat{A}_F(\omega-\Omega_F)M(\omega,z) L_W R(z) 
\label{eqq:FF}
\end{equation}
}
where $M(\omega,z)=exp[i\Delta\kappa_F(\omega) z_0+i\kappa(\omega)z]\exp[-\xi(\omega)^2]$ and $\Delta\kappa_F(\omega)=\kappa_{F}+\kappa(\Omega_F)+(\omega-\Omega_F)\kappa_{\Omega_F}'-\kappa(\omega)$.
The function $\xi(\omega)=\Delta\kappa_F(\omega)L_W/2$ rules the frequency response of the flash.
The ephemeral nature of the flash leads to the nearly staircase profile for the growth of the DW amplitude along the distance $z$
\begin{align}
R(\omega,z)=Erfi\left[\xi(\omega)+i\frac{z-z_0}{L_W} \right]-Erfi\left[\xi(\omega)-i\frac{z_0}{L_W} \right]
\label{eqq:RR}
\end{align}
where we used the complex error function $Erfi(y)=Erf(iy)/i$;  note that $R(0)=0$  for an input condition $F_F(0)=0$. From Eq. (\ref{eqq:FF}) we may observe that the phase matching condition $\Delta\kappa_F(\omega)=0$ induces a bandpass filtering behavior, with a bandwidth which is inversely proportional to the collision length $L_W$.
{\color{black} We verified numerically that the nonlinear dynamics of DW induced by a soliton collision  does not change significantly when including the Raman effect.}

\section{Numerical Simulations and discussion}
Let us compare DW generation by individual solitons with that resulting from soliton collisions. For that purpose, we perform full numerical simulations of the NLSE Eq. (\ref{eqq:nlse}) using the split-step Fourier method, and we develop analytical solutions of the DW generator Eq. (\ref{eqq:diflin}). In the numerics, we used  the exact profile of $\beta(\omega)$ as determined by a mode solver. 

\begin{figure}[!t]
\centering
\includegraphics[width=2.8in]{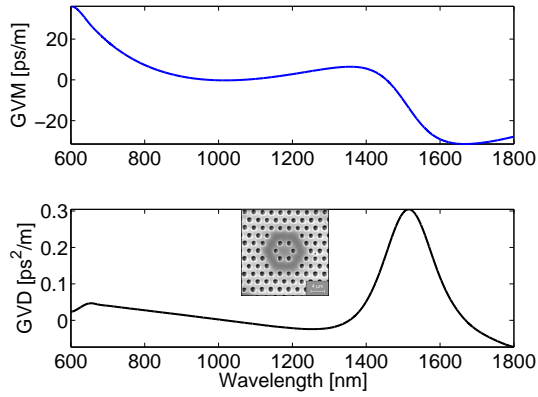}
\caption{Top: group velocity mismatch (GVM). Bottom: group velocity dispersion (GVD) of the fundamental even supermode of the dual-core MOF; the fiber cross-section is shown in the inset. }
\label{figuredisp}
\end{figure}

\begin{figure}[!t]
\centering
\includegraphics[width=3.0in]{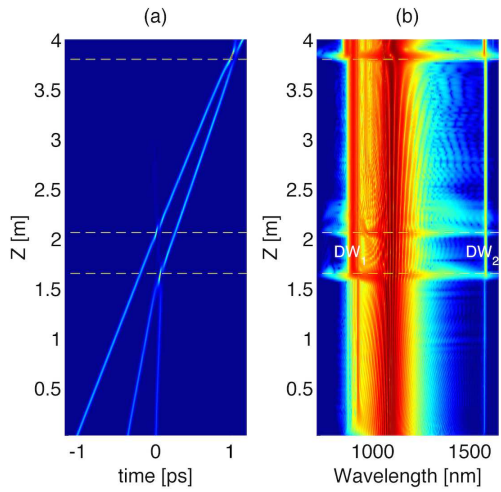}
\caption{Contour plots of optical power from three-soliton collision in dual-core PCF: (a) time-domain; (b) frequency-domain. The color scale of the spectrum is in dB.}
\label{figure5}
\end{figure}

In Fig. \ref{figuredisp} we show the dispersion law of the dual concentric core MOF we used as test bench for our analysis. We report in the inset the fiber cross section taken by a scanning electron microscope  (see \cite{us} for further details). 
Consider now the propagation and collision of two $35$ fs (full-width at half maximum) solitons, with 
$U_0$ centered at $\lambda_0=1064$ nm and energy of $29$ pJ and $U_1$ centered at $\lambda_1=1081.7$ nm 
and energy of $39$ pJ ($\gamma=21$ $ W^{-1}km^{-1}$).  
In numerical NLSE simulations we also included the presence of a third soliton at $\lambda_2=1100$ nm, so that we may infer how the DWs accumulate after two successive collisions.
We supposed three solitons with identical temporal duration, so that their peak powers vary in accordance with the different GVD values at each carrier wavelength. Proper initial time delays were added to the initial pulses, to observe the two collisions one after the other. Figure \ref{figure5} illustrates the history of the three-soliton collision process in the time and frequency domains, respectively. Analysis of the phase-matching condition shows that two separate resonant wavelengths (i.e., $\lambda_{DW_1}\simeq 900$ nm and $\lambda_{DW_2}\simeq 1587$ nm ) are leading to collision-induced DWs, which are indicated in Fig. \ref{figure5} as $DW_1$ and $DW_2$, respectively. 
As it can be seen, each collision results in a staircase growth of the energy in the DW spectra. However the two DWs exhibit quite different features: $DW_1$ acquires energy through spectral broadening, while $DW_2$ increases its peak intensity and keeps a relatively narrowband spectrum.
Figure \ref{figure5} also shows that each of the soliton-soliton collisions is strongly inelastic: after each collision, one of the solitons is nearly annihilated. 

\begin{figure}[!t]
\centering
\includegraphics[width=3.2in]{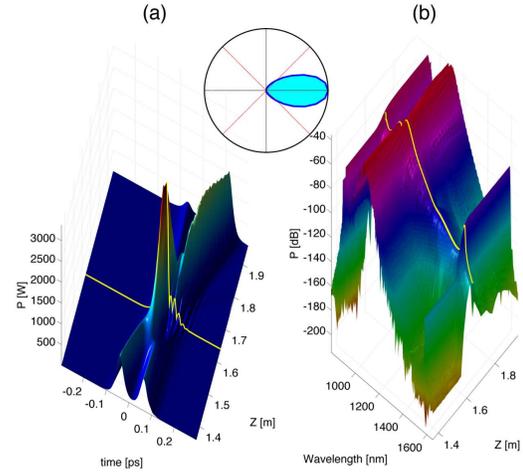}
\caption{Color online. Detailed view of optical power evolution during the first collision. The yellow curve highlights the flash pulse; (a) time-domain; (b) frequency-domain. The inset shows in polar coordinates and linear scale the intensity of $DW_2$ upon the input phase difference between the colliding solitons.}
\label{figure6}
\end{figure}

In Fig. \ref{figure6} we display the details of the collision of the two solitons at $\lambda_0$ and $\lambda_1$: Fig. \ref{figure6}(a) shows that for a short propagation distance the collision generates a narrow and intense flash, which later decays in two solitons of unequal amplitudes.
The yellow curve highlights the temporal and spectral profiles at the collision point: Fig. \ref{figure6}(b) clearly reveals the nearly stepwise growth of the $DW_2$ power at 1587 nm  ($0.25$ fJ after collision).
 Note that this is the energy contribution of a single collision: in a situation including 
thousands of interacting solitons  the overall DW energy may grow larger by several orders of magnitude, as numerically and experimentally demonstrated in Ref. \cite{us}.
The inset summarizes in a polar plot a series of numerical simulations of the same collision under different input soliton phases: the radius shows the intensity of $DW_2$ (linear scale) while the angle 
gives the initial phase difference between the two solitons. 
The maximum of DW emission is obtained for an in-phase collision. At the collision, the flash pulse spectrum 
gets broader and its spectral wings can be approximated by a $sech$ function which is nearly two times broader than the individual soliton spectra. 
The contribution to the DW brought by the collision vanishes when the two solitons are out of phase.

\begin{figure}[!t]
\centering
\includegraphics[width=3.2in]{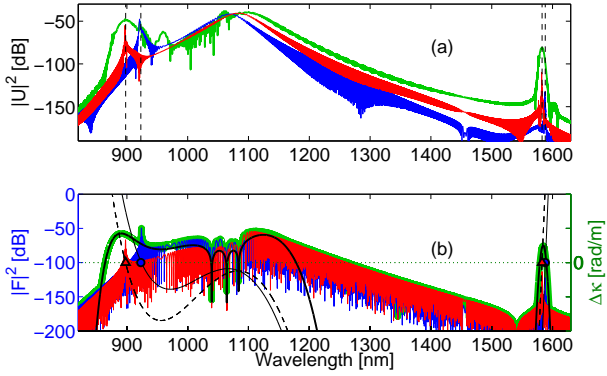}
\caption{Color online. (a) Numerical solution of the NLSE after 4 m of fiber: from two-soliton collision (green curve), and from individual solitons $U_0$ (blue), $U_1$ (red). (b) Blue and red curves show the analytical solutions of Eq. (\ref{eqq:s1}) for $S_{0}$ and $S_{1}$; the thick black curve shows the analytical DW spectrum of Eq. (\ref{eqq:FF}). 
The green curve combines the analytical solutions for two-soliton collision and for the two individual solitons. Thin black solid and thin black dashed curves show the phase mismatch $\Delta\kappa$ for soliton $U_0$ ($\Delta\kappa_0$) and $U_1$
($\Delta\kappa_1$); marks show corresponding phase matching points. }
\label{figure7}
\end{figure}

In Fig. \ref{figure7}(a) we compare the spectrum at the output of a 4 m long dual-core MOF which results after a two-soliton collision (green curve), with the spectra obtained when each of the two solitons propagates individually (blue and red curves for $U_0$ and $U_1$, respectively). Here the spectra are numerically computed from Eq. (\ref{eqq:nlse}). As it can be seen, a two-soliton collision brings about a largely broadened $DW_1$ spectrum around 900 nm, as well as more than 25 dB of power enhancement of the $DW_2$ narrow spectral peak.

To approximate the exact two-soliton collision, we took a flash pulse with peak power four times higher than a single soliton and with half time duration; the collision length $L_W$ was 5 cm. 
We assumed here $\Omega_F=\Omega$ as it is also confirmed by the numerical solution of Eq.(\ref{eqq:nlse}).
By using Eq. (\ref{eqq:FF}), one may thus analytically estimate the contribution of the flash pulse (i.e., of the soliton collision) to the overall DW spectrum, as it is shown in Fig. \ref{figure7}(b).
Here the black solid curve shows the solution of Eq. (\ref{eqq:FF}): the extra DW spectral components which appear in the numerical solution of Fig. \ref{figure7}(a) in the presence of the collision are indeed well reproduced by the estimate that is provided by the flash pulse contribution.  
In Fig. \ref{figure7}(b) we have also summed all solutions of the linearized equations for $S_{0}$, $S_{1}$, $S_{F}$ (green curve): this overall analytic DW spectrum reproduces most of the spectral features which are observed in the numerical solution of Fig. \ref{figure7}(a). For example, we may also recognize in Fig. \ref{figure7}(b) the sharp peak around $920$ nm which 
is generated by the soliton $U_0$.
In Fig. \ref{figure7}(b), the thin black curves show the wavelength dependence of the phase mismatch $\Delta\kappa$ associated with $U_0$ and $U_1$: positions where $\Delta\kappa=0$ identify the location of single soliton DW emission (see marks). 
\begin{figure}[!t]
\centering
\includegraphics[width=3.2in]{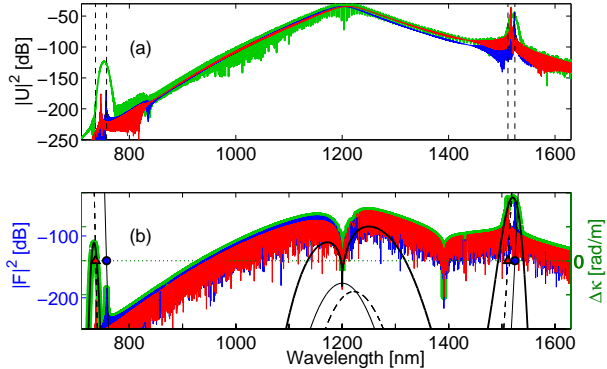}
\caption{Color online. Same as in Fig. \ref{figure7} but with soliton $U_0$ at 1200 nm and soliton $U_1$ at 1222 nm.
Here we  assumed $\Omega_F=0$ as the center wavelength of the flash is close to that of soliton $U_0$.}
\label{figureadd}
\end{figure}
 We show in Fig. \ref{figureadd} a different situation where the solitons $U_0$ and $U_1$ are centered at 
1200~nm and 1222~nm respectively and with the same pulsewidth. {\color{black} Once again, the agreement among numerics (Fig. \ref{figureadd}(a)) and analytical results (Fig. \ref{figureadd}(b)) is remarkable}. Note that in this case the collision brings 600 fJ to $DW_2$ that is
2400 times larger than that of Fig. \ref{figure7} for soliton energies four times bigger. The efficiency grows larger when the DW spectrum gets closer to the soliton's central wavelength.  A proper fiber design and choice of the central wavelengths 
can improve the intensity of the DW emission.

\section{Conclusion}
In conclusion, we have theoretically evaluated the effect of soliton-soliton collisions on the DW spectral amplitude and bandwidth.
Our analysis may provide a useful analytical guideline for {\color{black} predicting} the spectral location, bandwidth and energy of DWs generated in microstructured fibers via the temporal control of two distinct  trains of femtosecond pulses \cite{demircan}, thus widening
the range of supercontinuum sources  and spectroscopic instruments.  

\section*{Acknowledgment}
We acknowledge the support of the project Dat@Diag financed by OSEO France and by Horiba Medical, and of the Italian Ministry of University and Research (grant no. 2012BFNWZ2)

%\bibliographystyle{IEEEtran}

\end{document}